\documentclass[12pt]{article}
\usepackage{amsmath}
\usepackage{amssymb} 
\usepackage{bbm} 
\usepackage[small]{caption2} 
\usepackage{fleqn} 
\usepackage{graphicx} 
\usepackage{mathrsfs}   
\usepackage[small,loose]{subfigure}  
\usepackage{cite} 
\usepackage{color}

\newcommand{\eVdist}{\kern-0.06em}

\newcommand{\gev}{\:\text{Ge\eVdist V}}
\newcommand{\tev}{\:\text{Te\eVdist V}}

\DeclareMathOperator{\re}{Re}
\DeclareMathOperator{\im}{Im}

\newcommand{\Z}[1]{\ensuremath{\mathbbm{Z}_{#1}}} 

\newcommand{\hu}{\ensuremath{H_{u}}}
\newcommand{\hd}{\ensuremath{H_{d}}}
\newcommand{\qhu}{\ensuremath{q_{\hu}}}
\newcommand{\qhd}{\ensuremath{q_{\hd}}}
\newcommand{\singlet}{\ensuremath{S}}

\newcommand{\be}{\begin{equation}}
\newcommand{\ee}{\end{equation}}
\newcommand{\bea}{\begin{eqnarray}}
\newcommand{\eea}{\end{eqnarray}}

\allowdisplaybreaks[1]

\begin{document}

\begin{titlepage}

\vspace*{-3.0cm}
\begin{flushright}
OUTP-11-48P
\end{flushright}

\begin{center}
{\Large\bf
 The fine-tuning of the generalised NMSSM
}

\vspace{1cm}

\textbf{
Graham G.~Ross,
Kai Schmidt-Hoberg
}
\\[5mm]
\textit{\small
Rudolf Peierls Centre for Theoretical Physics, University of Oxford,\\
1 Keble Road, Oxford OX1 3NP, UK
}
\end{center}

\vspace{1cm}

\begin{abstract}
We determine the degree of fine-tuning needed in a generalised version of the NMSSM that follows from an underlying $\Z{4}$ or $\Z{8}$ $R$-symmetry.  We find that it is significantly less than is found in the MSSM or NMSSM and extends the range of Higgs mass that have acceptable fine-tuning. Remarkably the minimal fine-tuning is achieved for Higgs masses of around $130 \gev$. 
\end{abstract}

\end{titlepage}

\section{Introduction}

Supersymmetric extensions of the standard model (SM) offer a solution
to the hierarchy problem and can therefore generate a consistent Grand Unified or superstring unified model of the fundamental interactions. The remarkable success of the resultant prediction for gauge coupling unification has prompted  an enormous body of work on SUSY phenomenology. The most immediate prediction is that there should be a spectrum of supersymmetric states with masses of the order of the 
electroweak breaking scale. However, none of the superpartners have been discovered yet and this has given rise to the little hierarchy problem, namely the difficulty in explaining why SUSY states should be so much heavier than the $W$ and $Z$ bosons.

One may obtain a quantitative measure of the little hierarchy problem via a fine-tuning measure that quantifies the degree of cancellation between uncorrelated terms in the Lagrangian needed to achieve the mass splitting\cite{Ellis:1986yg,Barbieri:1987fn}. The degree of fine-tuning has been widely studied in the context of the minimal supersymmetric extension of the SM (MSSM). In the context of gravity mediation most of the low-fine-tuned areas of the model have already been ruled out by the non-observation of the SUSY partners and the recent LHC results leave only two significant regions to be checked. 
One of these is accessible to the LHC in its first $7\;\tev$ run with $1 \text{fb}^{-1}$ of data while the second requires the full $14\;\tev$ CM energy or improvement by one or two orders of magnitude sensitivity in the dark matter searches\cite{Cassel:2011tg}. For the case of gauge mediation the fine-tuning is worse and over all parameter space the fine-tuning required is greater than 1 part in 100 \cite{Abel:2009ve}.

Given this, it is of particular interest to ask whether non-minimal SUSY extensions can ameliorate the fine-tuning needed and, if so, to determine the changes in phenomenology that may be implied. It has been known for some time\cite{BasteroGil:2000bw} that the next-to-minimal SUSY extension that involves an additional SM singlet superfield, $S$, does reduce the fine-tuning, particularly at low tan$ \beta$, see also \cite{Ellwanger:2011mu} for a recent analysis. In practice the studies have almost all been done assuming the theory has a non-$R$ symmetry that restricts the couplings in the superpotential to a trilinear singlet term together with a singlet coupling to the Higgs superfields (NMSSM); for general reviews of the NMSSM see~\cite{Maniatis:2009re,Ellwanger:2009dp}. As a result the $\mu$ term is only generated when the scalar component of $S$ acquires a vev, solving the $\mu$ problem as this term is of the visible sector SUSY breaking scale. 

Recently an elegant alternative solution of the $\mu$ problem has been proposed \cite{Lee:2010gv,Lee:2011dya} that uses a discrete $R$-symmetry to forbid the bare $\mu$ term. A $\mu$ term of the correct magnitude is then generated by gravity mediation when SUSY and the $R$-symmetry are broken in a hidden sector.  Moreover the same symmetry forbids baryon- and lepton- number violating terms (apart from the Majorana mass term for right-handed neutrinos) up to and including the non-renormalisable terms of dimension 5 in the Lagrangian - a significant improvement on the usual $R$-parity of the MSSM.  

For the case that the MSSM is extended to include an additional singlet the same $R$-symmetry allows additional terms in the superpotential compared to the NMSSM.  Interestingly it has already been noted that the presence of such additional terms can further reduce the fine-tuning needed \cite{Dine:2007xi, Cassel:2009ps, Delgado:2010uj}.  In this paper we extend these analyses to estimate the fine-tuning needed over the full parameter space available with the additional superpotential terms and find a significant reduction in the fine-tuning, particularly for Higgs masses around $130\;\gev$.  

\section{Singlet extensions of the MSSM} 
The most general extension of the MSSM by a gauge singlet chiral superfield consistent with the SM gauge symmetry, the GNMSSM, has a superpotential of the form
\begin{equation}
 \mathcal{W} = \mathcal{W}_\text{Yukawa} + (\mu + \lambda S) H_u H_d + \xi S+ \frac{1}{2} \mu_s S^2 + \frac{1}{3}\kappa S^3
 \label{gen}
 \;
\end{equation}
where $\mathcal{W}_\text{Yukawa}$ is the MSSM superpotential generating the SM Yukawa couplings. Allowing for SUSY breaking the soft terms associated with the Higgs and singlet sector are
\begin{align}
 V_\text{soft} 
   &=  m_s^2 |s|^2 + m_{h_u}^2 |h_u|^2+ m_{h_d}^2 |h_d|^2 \nonumber \\
   &+ \left(b\mu \, h_u h_d + \lambda A_\lambda s h_u h_d + \frac{1}{3}\kappa A_\kappa s^3 + \frac{1}{2} b_s s^2  + \xi_s s + h.c.\right) \;.
\label{soft}
\end{align}
Here large letters refer to superfields while small letters refer to the corresponding scalar component.

The free parameters of the GNMSSM relevant to the scalar sector at the electroweak scale are given by $\lambda, A_\lambda, \kappa, A_\kappa, \mu, b\mu, \mu_s, b_s, \xi, \xi_s$ and $m_{h_u}^2, m_{h_d}^2, m_s^2$.
We can trade the three soft masses for the Higgs vevs $v, v_s$ and $\tan\beta$. 
As discussed in Section \ref{potential}, radiative corrections introduce a further dependence on the stop mass, $M_{\tilde{t}}$.
Fixing $v$ to give the observed gauge boson masses leave 13 independent parameters.

\subsection{The NMSSM}
The problem with the most general set of couplings appearing in eq.~\eqref{gen} is that the gauge symmetry and supersymmetry do not constrain the dimensionful terms to be small. However to be viable they should be of the order of the supersymmetry breaking scale or less; in the case of the $\mu$ term this is the ``$\mu$ problem'' that also applies to the MSSM. A simple solution is to require that the renormalisable couplings are constrained by a  $\Z{3}$ symmetry acting on the singlet chiral superfield and the Higgs superfields. Its superpotential has the restricted form 
\begin{equation}
 \mathcal{W}_\text{NMSSM} = \mathcal{W}_\text{Yukawa} + \lambda S H_u H_d  + \frac{1}{3}\kappa S^3 \;.
 \label{NMSSM}
\end{equation}
It is parameterised by a subset  of the parameters of the GNMSSM given by $\lambda, A_\lambda, \kappa, A_\kappa, $ $v_s, \tan\beta$ and $M_{\tilde{t}}$. In particular the bare mass terms are absent and an effective $\mu$ term is generated once the $s$ field acquires a vev. Allowing for the soft SUSY breaking terms the latter is of order the SUSY breaking scale as is required for a viable theory and so the structure provides a nice solution to the $\mu$ problem. 

To date most studies have concentrated on this version of a singlet extension of the MSSM and we shall refer to it as the NMSSM. Of course, like the MSSM, an additional $\Z{2}$ $R$-parity must be imposed to forbid the baryon- and lepton- number violating operators of dimension 3 and 4. Even with this there remain troublesome dimension 5  operators that, to inhibit proton decay, must be suppressed by a very large mass scale some $10^{7}$ times the Planck mass! Also, to avoid problems with stabilisation of the hierarchy \cite{Abel:1996cr} and domain walls \cite{Abel:1995wk} an additional $R$-symmetry must be imposed \cite{Abel:1996cr,Panagiotakopoulos:1998yw}.

\subsection{The GNMSSM}
As mentioned above, an alternative solution to the $\mu$ problem has recently been developed \cite{Lee:2010gv,Lee:2011dya} based on a discrete $R$-symmetry. In the context of the general singlet extension of the MSSM the symmetry also explains why the other dimensionful terms are also constrained to be of order the SUSY breaking scale as is necessary for a viable model. A further advantage of these symmetries is that they also forbid all baryon- and lepton- number violating terms up to and including dimension 5 terms in the Lagrangian. As a result there is no need to introduce a super-Planckian mass scale. Due to the $R$-symmetry, there is also no need to enlarge the symmetry to avoid problems with the stability of the hierarchy and domain walls \cite{Lee:2011dya}. 
 
 Two discrete $R$-symmetries were identified, $\Z{4}^{R}$ and $\Z{8}^{R}$, under which the fields transform as given in Table \ref{ZN}.
 \begin{table}[!h!]
\centerline{
\begin{tabular}{|c|c|c|c|c|c|}
\hline
 $M$ & $q_{\boldsymbol{10}}$ & $q_{\overline{\boldsymbol{5}}}$ & 
 	$\qhu$ & $\qhd$  
	& $q_{S}$ \\
\hline
 4 & 1 & 1 & 0 & 0  & 2 \\
 8 & 1 & 5 & 0 & 4 & 6 \\
\hline
\end{tabular}
}
\caption{Charge assignments under $\Z{M}^{R}$ for $M=4,8$. The labels $10$ and $\bar{5}$ refer to the $SU(5)$ matter content.}
\label{ZN}
\end{table}
Before SUSY/$R$-symmetry breaking the superpotential is of the NMSSM form in eq.\eqref{NMSSM}. However after supersymmetry breaking in a hidden sector with gravity mediation dimensionful terms are generated. With these the renormalisable terms in the superpotential take the form \cite{Lee:2011dya}
\begin{eqnarray}
 \mathcal{W}_{\Z{4}^{R}}
 & \sim &  \mathcal{W}_\text{NMSSM}+ m_{3/2}^2\, \singlet + m_{3/2}\, \singlet^{2} 
               + m_{3/2}\, \hu\, \hd \;,\\
   \mathcal{W}_{\Z{8}^{R}}
  & \sim &  \mathcal{W}_\text{NMSSM} + m_{3/2}^2\, \singlet 
\label{eq:WNMSSM1}
\end{eqnarray}
where the $\sim$ denotes that the dimensional terms are specified up to $\mathcal{O}(1)$ coefficients.

One may see that  $\mathcal{W}_{\Z{4}^{R}}$ is of the form of eq.\eqref{gen}, the GNMSSM superpotential,  but with the constraint that the scale of the dimensionful couplings is set by $m_{3/2}$, the scale of supersymmetry breaking in the visible sector. After eliminating the linear term in $S$ by a shift in its vev we obtain the general form of the superpotential given by 
\begin{equation}
 \mathcal{W}_\text{GNMSSM} = \mathcal{W}_\text{Yukawa} + (\mu + \lambda S) H_u H_d + \frac{1}{2} \mu_s S^2 + \frac{1}{3}\kappa S^3
 \label{GNMSSM}
 \;
\end{equation}
with $\mu\sim\mu_{s}\sim \mathcal{O}(m_{3/2})$.
We will refer to this model as the GNMSSM. In the case the underlying symmetry is $\Z{8}^{R}$ we obtain the same form for the superpotential but with the constraint $\mu/\mu_{s}=\lambda/(2 \,\kappa)$.

In \cite{Delgado:2010uj} a slightly restricted form of the GNMSSM was studied. Their model, referred to as the SMSSM, has a restricted set of the parameters appearing in eqs.~\eqref{gen} and \eqref{soft}, corresponding to setting $\kappa$ and $A_{\kappa}$ to zero.

\section{The scalar sector} 
\label{potential}
Fine-tuning refers to the difficulty in explaining why the electroweak breaking scale is so much less than the SUSY breaking scale. In order to provide a quantitative measure of fine-tuning it is necessary first to study the scalar potential that determines the electroweak breaking scale.
In addition to the terms which originate from the superpotential and the soft terms given above (cf.~eqs.~\eqref{gen} and \eqref{soft}),
we also have to take into account the $D$-term potential which is given by
\begin{equation}
 V_D = \frac{g_Y^2+g_2^2}{8}\left(|h_u|^2-|h_d|^2\right)^2  +\frac{g_2^2}{2}|h_d^\dagger h_u|^2 \;.
\end{equation}
In our convention the Higgs fields are expanded around the neutral real vevs as
\begin{equation}
 h_u^0 = v_u + \frac{1}{\sqrt{2}}\big(\re(h_u^0) + \text{i} \im(h_u^0)\big)
\end{equation}
etc.\ such that 
\begin{equation}
 v^2 = v_u^2 + v_d^2 \simeq (174 \gev)^2 \;.
\end{equation}
It is well known that radiative corrections significantly affect the Higgs potential.
For simplicity we only consider the leading one-loop contribution.
Following \cite{Cassel:2009ps} we add a term 
\begin{equation}
  \frac{g^2}{8} \delta |h_u|^4 \equiv \frac{g_Y^2+g_2^2}{8} \delta |h_u|^4
\end{equation}
to the Higgs potential, with
\begin{equation} 
\delta  =\frac{3\,h_{t}^{4}}{g^{2}\,\pi ^{2}\,} \ln \frac{M_{\tilde{t}
}}{m_{t}} = \frac{3 m_t^4}{2 \pi^2 M_Z^2 v^2 \sin^4\beta} \ln \frac{M_{\tilde{t}}}{m_t} \;.
\end{equation}
Here $M_{\tilde{t}}$ is the average stop mass while $m_t$ and $h_t$ are the top
mass and Yukawa coupling respectively.

The condition for the scalar potential to develop a minimum at $v_u,v_d,v_s$ can be written as 
\begin{align}
& m_{h_u}^2 + (\lambda v_s+\mu)^2 - \frac{1}{4} v^2 (g^2 - 4 \lambda^2) \cos^2\beta \nonumber \\
& \quad - (b\mu + \lambda (v_s (A_\lambda + v_s \kappa + \mu_s) + \xi)) \cot\beta + 
\frac{1}{4} g^2 v^2 (1 + \delta) \sin^2\beta =0  \:, \\
& m_{h_d}^2 + (\lambda v_s+\mu)^2 - \frac{1}{4} v^2 (g^2 - 4 \lambda^2) \sin^2\beta \nonumber \\
& \quad - (b\mu + \lambda (v_s (A_\lambda + v_s \kappa + \mu_s) + \xi)) \tan\beta + 
\frac{1}{4} g^2 v^2  \cos^2\beta =0 \; ,\\
& (v^2 \lambda \mu + \mu_s \xi - \xi_s
   v^2 \lambda (A_\lambda + \mu_s) \cos\beta \sin\beta) + (b_s + m_s^2 + v^2 \lambda^2 + \mu_s^2 
\nonumber \\
& \quad + 2 \kappa \xi - 
    v^2 \kappa \lambda \sin(2 \beta)) v_s + \kappa (A_\kappa + 
    3 \mu_s) v_s^2 + 2 \kappa^2 v_s^3=0 \;.
\end{align}

The mass matrix of the CP even Higgs fields in the basis ($h_u,h_d,s$) can be written as 
\begin{align}
\widetilde{M}_{11}^2 &= \cot (\beta ) (\lambda  (v_s(A_{\lambda}+v_s \kappa +\mu_s)+\xi )+b\mu)+M_Z^2 \sin ^2(\beta )(1+\delta) \:,\nonumber \\
\widetilde{M}_{22}^2 &= \tan (\beta ) (\lambda  (v_s(A_{\lambda}+v_s\kappa +\mu_s)+\xi )+b\mu)+M_Z^2 \cos ^2(\beta )\:,
\nonumber\\
\widetilde{M}_{33}^2 &=  v_s \kappa  (A_{\kappa}+4 v_s\kappa +3 \mu_s)+\frac{1}{2 v_s}\left(v^2 \lambda  (A_{\lambda}+\mu_s) \sin (2 \beta )-2 v^2 \lambda  \mu -2 \mu_s
   \xi-2\xi_s\right)\:,
\nonumber \\
\widetilde{M}_{12}^2 &=-\lambda  (v_s(A_{\lambda}+v_s\kappa 
+\mu_s)+\xi )-b\mu+ (v^2 \lambda ^2 -M_Z^2/2)\sin (2 \beta )\:,
\nonumber \\
\widetilde{M}_{13}^2 &=v \lambda  (2 \sin (\beta ) (v_s\lambda +\mu )- \cos (\beta ) (A_{\lambda}+2 v_s\kappa +\mu_s))\:,
\nonumber\\
\widetilde{M}_{23}^2 &=v \lambda (2 (v_s \lambda + \mu) \cos(\beta) - (A_{\lambda} + 
      2 \kappa v_s  + \mu_s) \sin(\beta))\:,
      \nonumber
\end{align}
where the soft masses $m_{h_u}^2, m_{h_d}^2$ and $m_s^2$ have been eliminated through the minimisation conditions.
Sometimes it is useful to consider the basis in which only one of the (doublet) Higgs $h$ obtains a vev.
This is done by a rotation of angle $\beta$ between the two doublet Higgs fields. In this basis $(h,H,s)$ the mass matrix then reads
\begin{align}
M_{11}^2 &= M_Z^2 (\cos^2(2\beta)+\delta \sin^4\beta) + \lambda ^2 v^2 \sin^2(2\beta) \;,  \nonumber \\
M_{22}^2 &=\frac{b\mu+ \lambda \xi + \lambda v_s(A_\lambda+v_s \kappa + \mu_s)}{\sin(\beta)\cos(\beta)} + (M_Z^2(1+\tfrac{\delta}{4})-\lambda^2 v^2)\sin^2(2\beta)\;,
\nonumber\\
M_{33}^2 &= v_s \kappa  (A_{\kappa}+4 v_s\kappa +3 \mu_s)+\frac{1}{2 v_s}\left(v^2 \lambda  (A_{\lambda}+\mu_s) \sin (2 \beta )-2 v^2 \lambda  \mu -2 \mu_s
   \xi - 2 \xi_s\right)\;,
   \nonumber \\
M_{12}^2 &=-\tfrac{1}{2}(M_Z^2-\lambda^2 v^2) \sin (4 \beta )+ \delta M_Z^2 \cos\beta \sin^3\beta \;, 
\nonumber\\
M_{13}^2 &= v \lambda  (2 v_s\lambda +2 \mu -\sin (2\beta )  (A_{\lambda}+2 v_s\kappa +\mu_s) )\;,
\nonumber\\
M_{23}^2 &=- v \lambda  \cos (2 \beta ) (A_\lambda +2v_s\kappa +\mu_s )\;.
\nonumber
\end{align}
The smallest eigenvalue is always bounded from above by the smallest diagonal entry.
In particular $M_{11}^2$ corresponds to the well-known upper bound on the lightest Higgs mass in the singlet extended MSSM. 
In comparison to the MSSM case there are additional terms which depend
on the trilinear coupling $\lambda$. One can see from the form of $M_{11}^{2}$ that the lightest Higgs mass can become larger for larger coupling $\lambda$.
However, the coupling $\lambda$ is bounded from above by the requirement of perturbativity
up to the GUT scale. For $\tan\beta \gtrsim 2$ this roughly corresponds to $\lambda \lesssim 0.7$, cf.~e.g.~\cite{Ellwanger:2009dp}.                                                                                                 
Furthermore, a large value of $\lambda$ generically leads to sizable mixing between the singlet and doublet Higgs fields, 
which decreases the lightest Higgs mass. 
This can be avoided by tuning $A_\lambda$ such that the matrix element $M_{13}$ and hence this mixing becomes zero.
Note, however, that starting from a specific high energy theory that determines the boundary values of the parameters it may not be possible to do this. 

Once we fix $\tan \beta$ the elements $M_{11}^2$ and $M_{12}^2$ cannot be varied independently.
However the remaining five independent elements can be varied independently by adjusting the free parameters.
The situation is the same in the usual NMSSM and for its generalised version.
In this sense the new parameters of the generalised NMSSM don't introduce additional freedom in the mass matrix.
Nevertheless the corresponding fine-tuning can be different in both cases and will be studied in the following.

\section{Fine-Tuning}

As introduced in \cite{Ellis:1986yg, Barbieri:1987fn}, a quantitative estimate of the the fine-tuning of the EW scale with respect to a set of independent parameters, $p$, is given by
\begin{equation} 
\Delta \equiv \max {\text{Abs}}\big[\Delta _{p}\big],\qquad \Delta _{p}\equiv \frac{\partial \ln
  v^{2}}{\partial \ln p} = \frac{p}{v^2}\frac{\partial v^2}{\partial p} \;.
\end{equation}
The quantity $\Delta^{-1}$ gives a measure of the accuracy to which independent parameters must be tuned to get the correct electroweak breaking scale. The independent parameters are usually taken to be those at a high energy scale determined by some underlying theory relevant at that scale. To determine them at the electroweak scale one has to compute their renormalisation group running. 

While the form of the GNMSSM superpotential is expected quite generically, the soft
supersymmetry breaking terms as well as the associated fine-tuning depend on the details of the supersymmetry breaking sector.
In what follows we choose a simple scheme for supersymmetry breaking, similar to the well-known CMSSM.
We take the high scale parameters to be a common gaugino mass, $m_{1/2}$, a common A-term, $A_0$, bilinear and soft bilinear terms for the doublet and singlet Higgs fields,
$\mu_0,\; b\mu_0, \;\mu_s^0,\; b_s^0$, and a common scalar mass $m_0$ for all scalar fields. The parameter set is completed by the Yukawa couplings $\lambda_0$ and $\kappa_0$.\footnote{Using the modified
definition for fine-tuning~\cite{Ciafaloni:1996zh}, appropriate for measured parameters, the contribution from the top Yukawa coupling is sub-dominant~\cite{Cassel:2010px}.}
We therefore have 9 free parameters in the GNMSSM case.
To obtain the relations between electroweak and the high scale parameters we solve the one-loop RGEs
\cite{Ellwanger:2009dp} numerically for different GUT scale boundary conditions and 
use it to interpolate over the full parameter space. The corresponding numerical relations we use can be found in the Appendix. 
In the case of the NMSSM it is well known that with universal $m_0$, $m_{1/2}$ and $A_0$ the Higgs mass cannot be enhanced with respect
to the MSSM case and we will therefore not consider the NMSSM case any further. A more general structure of the SUSY breaking terms
which also allows an enhancement of the Higgs mass in the NMSSM will be presented elsewhere.  

\subsection{Fine-tuning scan}

\begin{figure}[!h!]
\centerline{
\includegraphics[width=7.5cm]{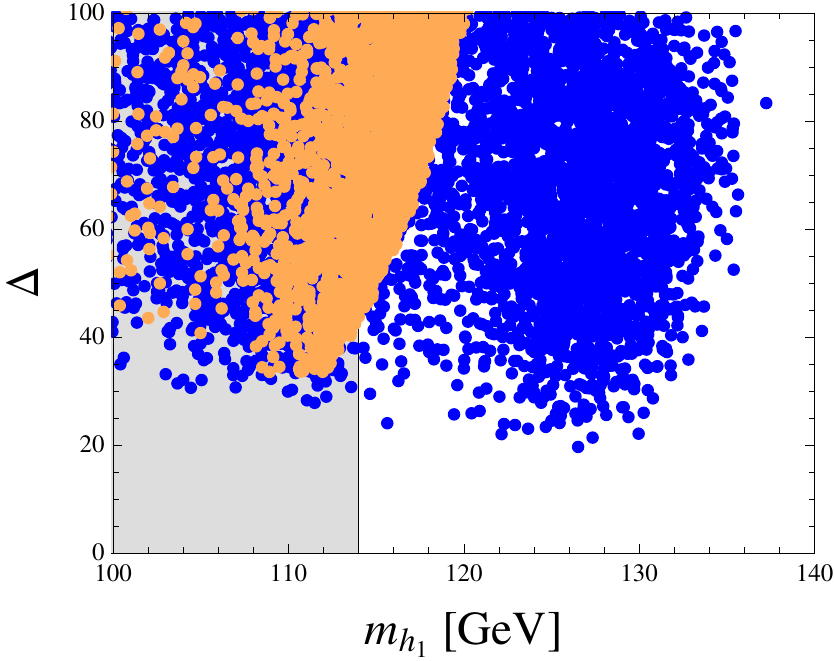}
\includegraphics[width=7.5cm]{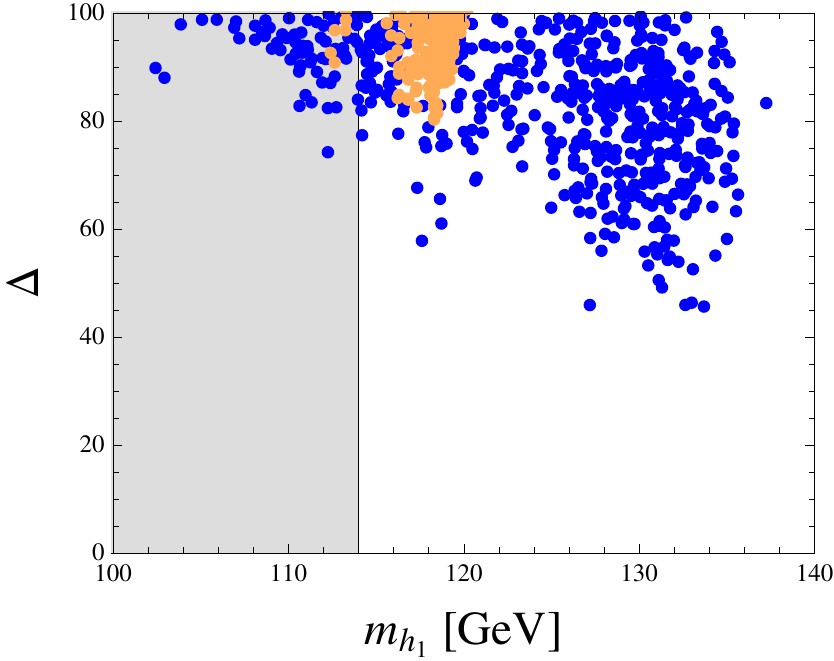}}
\caption{Comparison of fine-tuning in the MSSM (orange) and GNMSSM (blue) for a minimal unified gaugino mass  $m_{1/2} > 300 \,(500) \gev $. For the GNMSSM the lowest fine-tuning is at low $\tan\beta=\mathcal{O}(2)$ while in the MSSM it is at higher $\tan\beta=\mathcal{O}(10)$.  Remarkably the lowest fine-tuning is achieved for rather large Higgs masses, $m_{h_1} \sim 130 \gev$.}
\label{fig:fine}
\end{figure}

At the low scale three input parameters may be traded for $v,v_s$ and $\tan\beta$ using the minimisation conditions.
We choose these parameters to be $\mu_0, b\mu_0, b_s^0$ for the GNMSSM and $\mu_0, b\mu_0$ for the MSSM. 
We performed a scan over the remaining parameter space of the MSSM and the GNMSSM to compare their corresponding fine-tuning. 
The result of these scans is shown in Figure~\ref{fig:fine}.

In both cases we scanned over $300 \, (500) \le m_{1/2} \le 1000 \gev$, $0 \le m_0 \le 1000 \gev$, $0 \le A_0 \le 1000 \gev$,
and $2 \le \tan\beta \le 10$. 
In addition, in the GNMSSM we scanned over
$0.5 \le \lambda_0 \le 2$, $0 \le \kappa_0 \le 0.5$, $-1000\gev \le v_s \le 1000 \gev$, and $-5000 \gev \le \mu_s^0 \le 5000 \gev$.
Note that we restricted the effective $\mu$-term to $\mu_\text{eff} =(\mu +\lambda v_s)  \ge 100\gev$ 
to ensure that the charginos are above the experimental limit.
The other superparticle masses are predominantly determined by the universal gaugino mass $m_{1/2}$.
As a rough estimate for the current bounds from the LHC we impose a lower limit of $m_{1/2} > 300 \,  \gev $ and to demonstrate how things will change if the bounds improve we also show the case  $m_{1/2} > 500 \,  \gev $.
A more detailed numerical study of these constraints will be presented elsewhere.

It may be seen that the fine-tuning in the GNMSSM is reduced relative to the MSSM particularly at larger Higgs masses. 
Remarkably, the lowest fine-tuning is achieved for Higgs masses of $m_{h_1} \sim 130 \, \gev$. 
Comparing the cases for $m_{1/2}=300 \, \gev$ and $m_{1/2}=500 \, \gev$ it may be seen that although the overall fine-tuning increases with $m_{1/2}$, the Higgs mass corresponding to the smallest fine-tuning is almost unchanged.

\begin{figure}[!h!]
\centerline{
\includegraphics[width=7.5cm]{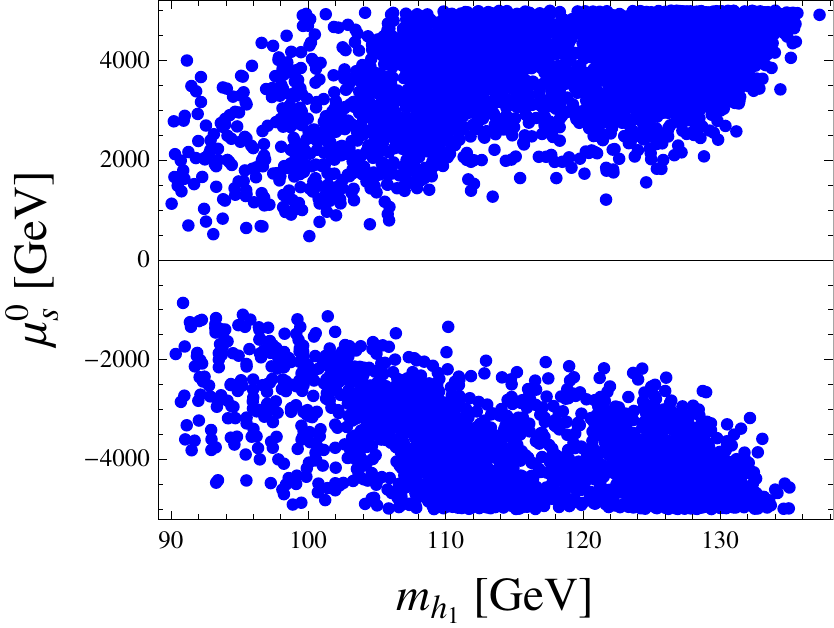}}
\caption{The lightest Higgs mass versus the high-scale singlet mass parameter $\mu_s^0$ for points with fine-tuning $\Delta <100$.}
\label{fig:mus}
\end{figure}

With the given universal soft parameters at the GUT scale we don't find any viable points with large Higgs masses and small values
of the singlet mass parameter $\mu_s^0$. This can also be seen in Figure~\ref{fig:mus}, where we show the lightest Higgs mass versus
$\mu_s^0$ for points which have a fine-tuning below 100. 
This means that the corresponding singlet states will be rather heavy leading to a phenomenology very similar to the MSSM, but with a smaller fine-tuning and a heavier Higgs than would be expected from the masses of the superpartners. It also implies that the slightly more `NMSSM-like' version following from an underlying $\Z{8}^R$ symmetry will have
somewhat larger fine-tuning than the more general $\Z{4}^R$ case, because the relation $\mu/\mu_{s}=\lambda/(2 \,\kappa)$ implies a $\mu$ term which is of similar size as $\mu_s$.
Note that both these conclusions crucially depend on the assumptions we have made about the supersymmetry breaking terms
and could be very different in a more general setup. In fact it is known that in the NMSSM with more general SUSY
breaking terms the singlet states can be light leading to interesting implications for phenomenology. We will consider these
more general SUSY breaking terms for the GNMSSM in a future work.
 
\subsubsection{Origin of the reduction in fine-tuning}
There are two main reasons for this reduction in fine-tuning. The first, which applies to both the NMSSM and GNMSSM, is that there is a contribution, $ \lambda ^2 v^2 \sin^2(2\beta)$, to the matrix element, $M_{11}^{2}$, that sets the upper bound on the lightest Higgs mass. This has the general effect of shifting the fine-tuning envelope in Figure \ref{fig:fine} to the right; how much depends on whether it is possible to saturate the bound with parameters maintaining low-fine-tuning. Due to its additional parameters this is easier in the GNMSSM. Note that the effect is significant only for low $\tan\beta \, (\tan\beta\sim 2).$ 

The second, which applies mainly to the GNMSSM, is the additional stabilising term in the Higgs potential that the additional singlet coupling generates and reduces the relative magnitude of the Higgs vevs to the SUSY breaking scale. This is most clearly seen in the large $\mu_{s}$ limit where one can integrate out the $S$ superfield at the supersymmetric level; this limit is applicable only to the GNMSSM and not to the NMSSM.  In this limit one obtains the superpotential term
$\lambda^{2} (H_{u}H_{d})^{2}/\mu_{s}.$ The leading contribution to the Higgs potential coming from this term is given by

\begin{equation}
V \simeq 4 \lambda^{2} (\lambda v_s + \mu) h_{u}h_{d}(|h_{u}|^{2}+|h_{d}|^{2})/\mu_{s}\;.
\end{equation}

This is proportional to $\sin 2\beta$ and so is peaked at low $\tan\beta$ although not as sharply as for the additional Higgs mass contribution.  The presence of this term was found \cite{Cassel:2009ps} to significantly reduce the fine-tuning relative to the NMSSM and explains the origin of the difference seen in Figure~\ref{fig:fine} for the case $\mu_{s}$ is large. For lower values of $\mu_{s}$ the singlet coupling generates stabilising terms in both the NMSSM and the GNMSSM although they are likely to be more important in the GNMSSM case.

\section{Summary and Conclusions}

Given the enthusiasm for SUSY as a solution to the hierarchy problem there has been surprisingly little attention paid to the general version of the NMSSM. This has been due to the apparent un-naturalness of the model which has additional mass terms in the Lagrangian that must be no larger than the SUSY breaking scale in the visible sector. However the recent discovery that simple discrete $R$-symmetries can naturally achieve this has made the GNMSSM as natural as the NMSSM. Indeed the fact that the symmetry also eliminates the dangerous dimension three, four {\it and} five baryon- and lepton-number violating terms in the Lagrangian and avoids destabilising tadpoles and domain  wall problems renders it a more promising starting point than the NMSSM. 

In this paper we have determined the fine-tuning in the GNMSSM for the case of gravity mediated SUSY breaking.
We found that the fine-tuning is significantly reduced compared to the MSSM and in particular 
significantly extends the range of Higgs mass that have acceptable fine-tuning.
Interestingly the region with smallest fine-tuning corresponds to Higgs masses of $\sim 130\gev$.

For the universal boundary conditions we consider the singlet mass scale entering in the superpotential was typically rather large compared to that of the other dimensionful parameters, implying that the phenomenology is very similar to that of the MSSM, although the fine-tuning can still be significantly reduced.

Our analysis of the GNMSSM has been somewhat rudimentary, relying on analytic approximations for the RGE running of parameters. Work on a more complete numerical analysis is in the planning stage and we hope to be able to present a more detailed study of the fine-tuning and associated collider phenomenology in the future. 

\section*{Acknowledgements}
We would like to thank Mads Frandsen and Dumitru Ghilencea for useful discussions. The research presented here was partially supported by the EU ITN grant UNILHC 237920 (Unification in the LHC era) and the ERC Advanced Grant BSMOXFORD 228169.

\appendix
\section{RGE running}
The independent high-scale parameters are given by $m_{1/2},A_0,\mu_0, b\mu_0, \mu_s^0, b_s^0,m_0, m_{h_u}^0, m_{h_d}^0$, $\lambda_0$ and $\kappa_0$.
Using the one-loop RGEs the relation between electroweak and GUT scale parameters can, for the low tan$\beta$ region of interest,  be approximately written as
\begin{eqnarray}
m_{h_u}^2   &\simeq & -0.10 \, A_0^2 - 0.77 \, m_0^2 + 0.39 \, A_0 \, m_{1/2} - 2.57 \, m_{1/2}^2 +  0.58 \, (m_{h_u}^0)^2 \nonumber \\
            && + \lambda_0 \, \left[ 0.21 \, m_{1/2}^2 - 0.08 \, (m_{h_d}^0)^2\right] \\
 m_{h_d}^2  &\simeq &  0.44 \, m_{1/2}^2 + 0.94 \, (m_{h_d}^0)^2 - \lambda_0 \, \left[0.07 \, m_0^2 + 0.11 \, (m_{h_d}^0)^2 + 0.10 \, (m_{h_u}^0)^2 \right] \\
\mu         &\simeq & 0.97 \, \mu_0 - 0.18 \, \lambda_0 \, \mu_0  \\
b\mu        &\simeq &   0.93 \, b\mu_0 - 0.40 \, A_0 \,\mu_0 + 0.22 \, m_{1/2} \, \mu_0 + \lambda_0 \, \left[0.07 \, A_0 \, \mu_0 - 0.38 \, b\mu_0 \right] \\
m_s^2       &\simeq & 0.90 \, m_0^2 +\lambda_0 \, \left[0.23 \,(m_{h_d}^0)^2- 0.19 \,(m_{h_u}^0)^2- 0.10 \,m_0^2\right] \nonumber \\
            &&      - 0.45 \, \kappa^0 \, m_0^2  + 0.14 \, \lambda_0 \, \kappa_0 \, m_0^2 \\
\mu_s       &\simeq & 0.97 \, \mu_s^0 - 0.36 \, \lambda_0 \, \mu_s^0 - 0.30 \, \kappa_0 \, \mu_s^0  +   0.12  \, \lambda_0 \, \kappa_0 \, \mu_s^0 \\
b_s         &\simeq& 0.93 \, b_s^0 - 0.11 \, A_0 \, \mu_s^0-\lambda_0 \, \left[0.298 \, b_s^0 + 0.10 \, A_0 \, \mu_s^0 \right] \nonumber \\
            &&       - \kappa_0 \, \left[0.38 \, b_s^0 + 0.05 \, A_0 \, \mu_s^0\right] + \lambda_0 \, \kappa_0 \, \left[0.17 \, b_s^0-  0.07 \, b\mu_0 \right] \\
\lambda     &\simeq& 0.78 \, \lambda_0 - 0.21 \, \lambda_0^2 \\ 
\kappa      &\simeq&  0.58 \, \kappa_0 - 0.16 \, \kappa_0^2\\
A_{\lambda} &\simeq& 0.54 \, A_0 + 0.23 \, m_{1/2}- 0.31 \, \lambda_0 \, A_0   - 0.16 \, \kappa_0 \, A_0 \\
A_{\kappa}  &\simeq&  0.84 \, A_0 +\lambda_0 \, \left[0.10 \, m_{1/2} - 0.52 \, A_0\right] - 0.42 \, \kappa_0 \, A_0 + 0.21 \, \lambda_0 \, \kappa_0 \, A_0 \\ 
m_{t_L}^2   &\simeq& 0.74 \, m_0^2 + 0.14 \, A_0 \, m_{1/2} + 5.24 \, m_{1/2}^2 - 0.14 \, (m_{h_u}^0)^2 + 0.08 \, \lambda_0 \, m_{1/2}^2 \\
m_{t_R}^2   &\simeq&  0.53 \, m_0^2 + 0.25 \, A_0 \, m_{1/2} + 3.90 \, m_{1/2}^2 - 0.20 \, (m_{h_u}^0)^2 + 0.13 \, \lambda_0 \, m_{1/2}^2 \\
A_t         &\simeq& 0.24 \,A_0 - 2.14\, m_{1/2} - 0.12 \, \lambda_0 \, m_{1/2}
\end{eqnarray}

\bibliography{NMSSM}
\bibliographystyle{ArXiv}

\end{document}